# Criticality estimation of IT business functions with the Business Continuity Testing Points method for implementing effective recovery exercises of crisis scenarios

Athanasios Podaras[1] and Tomas Žižka[2]

[1,2] Department of Informatics, Technical University of Liberec, Liberec, Czech Republic

**Abstract**
The primary goal of the present paper is the introduction of a new approach of defining IT unit business functions' exact criticality levels and respectively categorize them to the appropriate recovery tests, prior to their thorough documentation which includes actual desired recovery timeframes. The method is entitled as Business Continuity Testing Points and it is based on the concept of Use Case Points, a fundamental project estimation tool utilized for sizing of object-oriented system development. The aim of the contribution is to ameliorate the existing manual way of determining recovery time of IT business functions that is based exclusively on experience of IT personnel, by introducing a calculation method of multiple factors that can negatively affect the recovery process. The elimination of damage as a result of tested immediate response action in a crisis situation that disrupts core IT operations constitutes the aimed advantage of the proposed contribution.

## 1. Introduction

The present enterprise operations are fully dependent on information technologies, complex software applications and multiple corresponding business functions. Thus, one of the most important, and nowadays obligatory, tasks of the modern enterprises and organizations is the development and the establishment of an efficient and effective *Business Continuity Management* [1]. It is by many experts considered to be one of the key Areas of ICT Competencies [2]. Its imperative establishment in terms of enterprise operational policy and strategy stems from multiple and various unexpected and forecasted physical, human or even technical threats that many organizations have experienced within the recent years.

Immediate system recovery for minimization of operational and financial damage stems from a cautiously documented Business Continuity Plan (BCP) [3] [4]. The BCP specifies the methodologies and procedures required to backup and recover every functional units of the business [5]. Thus regular testing in a simulative but as real as possible mode of the most critical business functions should be scheduled for controlling the stable functionality of IT business functions in crisis situations. Criticality estimation of business functions and corresponding applications is a vital task to be solved by IT managers when planning business continuity testing exercises that concern their unit. The organization and implementation of successful business recovery tests presuppose the creation of a detailed and accurate documentation of the critical business functions and corresponding applications to the *Business Impact Analysis* (BIA) template [3] [4].

Primary discussed and analyzed topics of the present work are all placed within the context of *testing important IT business functions' rapid recovery according to a hypothetical simple, average or complex testing crisis scenario decided on the basis of corresponding process criticality level.*

The author's attempt is to develop and propose a standard methodology for defining the most crucial IT business functions and processes in order to thoroughly document them in the Business Impact Analysis document and subsequently propose the appropriate recovery test type after calculating and defining exactly their criticality level. The determination of this level is based on consideration of corresponding impact value level [6] and additionally the minimum of business functions which must be up and running to ensure the company's basic operation in emergency events. The applied theory behind the construction of the new approach is the Use Case Points method [7]. Use Case model is one of the most tested in practice methodologies for defining user requirements [8]. Additionally, the Business Continuity Management document core characteristic is that it constitutes a requirements' document as well, though from the scope of business function testing procedure within a crisis situation. Required task by the organizations is the recovery of a business function within the desired Recovery Time Objective (RTO) or even the Maximum Acceptable Outage

[4] [3], as they are determined by the Business Continuity team and documented in the BIA template.

Thus, by implementing the Use Case concept can help IT managers perform accurate and objective business function criticality scoring and respectively determine the testing approach of each function, according to the BIA process documented data. The scoring of function criticality is applied according to the rules of the Use Case Points methodology by considering multiple factors, both technical and environmental which ultimately affect the business function recovery process and the timeframe that it can be accomplished. The estimated effort required to "bring back to life" a vital IT business function will comprise of the key indicator of the corresponding testing approach intended to be applied to the specific function.

## 2. Testing critical IT business functions within a defined enterprise Business Continuity Management policy

From the point of view of scientific literature and industrial practice BC addresses questions of how to handle risk issues in the case that critical business processes of an organization fails [9]. Historically, BC addressed IT processes, later on, business processes came up as the final purpose of their supporting IT processes [10] [1].

The importance of Business Continuity Testing is outlined and thoroughly analyzed by many experts. Precisely it is stated that organizing regular exercises such as desktop and simulation drills is the only way to discover gaps and address them [11].

The current paper focuses on the IT department's successful documentation and testing of the most critical functions and processes; hardware and software should support critical business functions, so the IT functions, in large part, will be driven by all the other departments. HR might say "we have to have our payroll application"; marketing might say "without our CRM system, we can't sell any products"; manufacturing might say "without our automated inventory management system, we can't even begin to make anything." Therefore, the IT department's critical business functions are driven externally, to a large degree [12]. However, the successful IT business continuity management policy should focus to the immediate recovery of the indeed most important operations of the enterprise, defined by the ISO 22301:2012 as Minimum Business Continuity Objective, briefly stated as MBCO [4].

Creating a continuity plan is a long-term process and companies should review the existing documentation as an ongoing project [13] [14]. The actual purpose of testing is to achieve organizational acceptance that the solution satisfies the recovery requirements. Plans may fail to meet expectations due to insufficient or inaccurate recovery requirements, solution design flaws or solution implementation errors [3]. The differentiation of critical (urgent) and non-critical (non-urgent) organization functions/activities is the core task of B.I.A. Critical functions are those whose disruption is regarded as unacceptable. Acceptability is mainly affected by the cost of recovery solutions. A function may also be considered critical in the case that it is imperative due to specific law.

### 2.1 Exercise Categories

The exercises as they were defined by the Business Standards Institute are divided into 3 basic categories [3]:

**Tabletop Exercises:** they typically involve a small number of people and participants, who work through a simple scenario, discuss specific aspects of the plan and only a few hours are consumed.

**Medium Exercises:** conducted within a "Virtual World" and bring together several departments, teams or disciplines.

**Complex Exercises:** also occur within a virtual world, but maximum realism is essential and duration is unknown.

The results of insufficient and poor testing of software applications are known and obvious within the enterprise environment. Test engineering is seldom planned for in most organizations and as a result, products enter the market insufficiently tested. Negative customer reactions and damage to the corporate image is the natural consequence [15]. Similarly, the test engineering process for critical business functions from the business continuity management standpoint is essential, since the negative enterprise effects caused by unsuccessful response to a real crisis event will be an established fact. Consequently, according to the above statement all business functions should be tested regularly so that all involved staff will be prepared for a real crisis event. The idea behind the proposed contribution is that test success is based on mapping of each IT business function to the most suitable of the above analyzed exercise levels after determination of its impact value level. The way that the mapping is performed is depicted at the following section (see Table8).

### 2.2 Impact Value Levels of IT business functions

Darril Gibson [6] indicates the impact value level of each business function according to its accepted downtime period without causing negative effects to the enterprise or the organization. The four levels of impact value are:

**Level 1:** business functions must be available during all business hours. Online systems must be available 24 hours per day and 7 days per week.
Maximum Acceptable Outage (MAO) = 2 hours
Recovery Time Objective (RTO) < 2 hours

**Level 2:** business processes can survive without the business function for a short amount of time.
Maximum Acceptable Outage (MAO) = 24 hours (1 day)
Recovery Time Objective (RTO) < 24 hours

**Level 3:** business processes can survive without the business function for one or more days.
Maximum Acceptable Outage (MAO) = 72 hours (3 days)
Recovery Time Objective (RTO) < 72 hours

**Level 4:** business processes can survive without the business function for extended periods.
Maximum Acceptable Outage (MAO) = 168 hours (1 week)
Recovery Time Objective (RTO) < 168 hours

The above mentioned levels will be applied to the new proposed Business Continuity Testing Points methodology so that the responsible IT manager of a specific business function will be able to classify it to the appropriate exercise category.

## 3. The Use Case Points

Use Case points method [16] [17] [18] [15] is especially valuable in the context of early size measurement and effort estimation, because it employs use cases as an input [19], Use cases, proposed by Jacobson [20] [21], are a popular form of representing functional requirements. Moreover, according to the survey conducted by Neill and Laplante in 2003 [22], 50% of projects have their functional requirements presented as scenarios or use cases. They are also available in the early stages of soft-ware development. Thus, due to above stated important feature of functional requirements documentation and taking into consideration that the Business Continuity strategy is based on business functions recovery action requirements, Use Case method can comprise of a new approach to defining exact documentation in Business Impact Analysis documents and furthermore an assisting tool of estimating precisely the Recovery Time Objective and Maximum Acceptable Outage. As it was already mentioned the reason and the need for introducing the new method is to avoid the manual vague estimation of these values based solely on IT manager's practical experience. Before analyzing the new proposed model of Business Continuity Testing Points, a brief reference to the Use Case Points method is required.

### 3.1 Classifying Actors and Use Cases

The primary step of the Use Case Points procedure is to classify and calculate the Actors' weights and the weight of a Use Case. Classification method with regard to complexity degree of both Actors (see Table1) and Use Cases (see Table2) are listed respectively.

Table 1: Actor's Classification

| *Actor Type* | *Weighting Factor* |
|---|---|
| Simple | 1 |
| Average | 2 |
| Complex | 3 |

Simple Actor [7] [23] [24] represents another system with a defined Application Programming Interface, API, an average actor is another system interacting through a protocol such as TCP/IP, and a complex actor may be a person interacting through a GUI or a Web page.

By counting the number of actors of each kind (complex, average or simple), multiplying each total by its weighting factor and finally adding up the products we calculate the the total Unadjusted Actor Weights, briefly mentioned as UAW. The result of the calculation is provided by Eq. (1):

$$UAW = \sum_{i=1}^{n} A_i * W_i \qquad (1)$$

where n= Number of Actors, A= Actor, W= Actor's Weighting Factor.

In a similar way the Unadjusted Use Case Weight (UUCW) value is calculated by multiplying number of each use case category by the corresponding weighting factor, and the products are added up according to Eq. (2),

$$UUCW = \sum_{i=1}^{n} U_i * W_i \qquad (2)$$

where n= Number of Use Cases, U= Given Use Case, W= Use Cases weighting factor

The UAW is added to the UUCW to get the unadjusted use case points from Eq. (3)

$$UUCP = UAW + UUCW \qquad (3)$$

Table 2: Use Case Classification

| Use Case Type | No of Transactions | Weighting Factor |
|---|---|---|
| Simple | <=3 | 1 |
| Average | 4 – 7 | 2 |
| Complex | >7 | 3 |

3.2 Evaluation of Technical and Environmental Factors

In Use Case Points methodology apart from the computation of UUCP value, various Technical (see Table3) and Environmental Factors (see Table4) are considered and computed with respect to Software Application complexity.

After their computation the Adjusted Use Case Points (UPC) are calculated with the help of a special equation, in which Unadjusted Use Case Points value, Technical Complexity Factors (TCP) value and Environmental Factors (EF) value are multiplied.

Table 3: Technical Complexity Factors in Use Case Points

| Factor | Description | Weight |
|---|---|---|
| T1 | Distributed System | 2 |
| T2 | Response adjectives | 2 |
| T3 | End – User efficiency | 1 |
| T4 | Complex Processing | 1 |
| T5 | Reusable Code | 1 |
| T6 | Easy to install | 0.5 |
| T7 | Easy to Use | 0.5 |
| T8 | Portable | 2 |
| T9 | Easy to change | 1 |
| T10 | Concurrent | 1 |
| T11 | Security features | 1 |
| T12 | Access for third parties | 1 |
| T13 | Special Training Required | 1 |

Table 4: Environmental Factors in Use Case Points

| Factor | Description | Weight |
|---|---|---|
| F1 | Familiar with RUP | 1.5 |
| F2 | Application Experience | 0.5 |
| F3 | Object – Oriented experience | 1 |
| F4 | Lead Analyst capability | 0.5 |
| F5 | Motivation | 1 |
| F6 | Stable requirements | 2 |
| F7 | Part – time workers | -1 |
| F8 | Difficult programming language | 2 |

The formula applied for calculating Technical Complexity Factor (TCF), is provided by Eq (4):

$$TCF = 0.6 + (0.001 * TFactor) \qquad (4)$$

after multiplying the value of each Technical Factor (see Table3) by its corresponding weight and then adding all these numbers to get the sum called TFactor.

In the same way Eq. (5) is applied for calculating Environmental Factor (EF):

$$EF = 1.4 + (-0.03 * EFACTOR) \qquad (5)$$

after multiplying the value of each Environmental Factor (see Table4) by its corresponding weight and then adding all these numbers to get the sum called EFactor.

The final calculation of the Adjusted Use Case Points (UPC) is provided by Eq. (6):

$$UPC = UUCP * TFC * EF \qquad (6)$$

The estimation effort is the final part of the Use Case Points method. By multiplying the specific value (man-hours) by the UCP, estimated effort can be obtained [23] [25]. A factor of 20 man-hours per UCP for a project is suggested by Karner [7].

# 4. The Business Continuity Testing Points Method (BCTP)

4.1 Motivation, utilized theory and algorithmic process behind the BCTP methodology

The basic motivation for the construction of the BCTP approach as to defining exact exercise category for each IT function and enabling the implementation of effective recovery tests according to the defined RTO and MAO timeframes of business impact analysis documentation, has been the elimination of erroneous BIA documentation of high priority functions and processes which usually leads to poor testing implementation. As it was already mentioned, the basic reason is that RTO and MAO timeframes documentation in a BIA template is based on employers' everyday operational experience within the organization.

The following step was the selection of the appropriate existing and proved by practical implementation theory in order to derive the new model. With the above stated assumption, since the BIA documentation constitutes a basic element of the general Business Continuity Plan, which is considered to be a requirements' document from system recovery standpoint, a standard requirement analysis and practically implemented methodology such as Use Case Points approach was required as a driving method for the construction of the Business Continuity Testing Points approach. Moreover the testing goal is to satisfy organizational acceptance that the solution meets the recovery requirements.

As in the case of the Use Case Points methodology, the proposed model's construction is separated in 2 parts. The first part constitutes the mapping of Actors and Use Cases for calculation of Unadjusted Use Case Points, to the Actor Types 1 (Human Level) and 2 (Application Level) and Business Processes, instead of Use Cases, in order to calculate Unadjusted Business Function Testing Points (UBFTP). The idea is that in case of obvious and simple business function with low score of unadjusted points there will be no need for further analysis, and thus direct decision about impact value level and exercise category will be made in the final 2$^{nd}$ part. It is assumed that this simple approach concerns only business functions which are not included in MBCO and the corresponding exercise category planned will be either tabletop or medium. Exercise category is based on the derived number of unadjusted points and impact value level (3 or 4).

The second part constitutes a process intended for more complicated IT business functions for which complex exercises should be planned. Thus, exact impact value level should be decided first. The idea is the utilization of technical recovery, environmental recovery and also unexpected recovery factors which will enable better understanding of the exact impact value level (1 or 2) of the business function. The term *Unexpected Recovery Factors* is a new additional factor category included in the proposed model since Business Continuity Tests should be characterized by hypothetical unexpected conditions that will could prolong the recovery process.

Consequently the algorithmic steps to be implemented in terms of classifying IT business functions to the appropriate exercise category defined by the Business Standard Institution are the following:

*Step 1*: Defining Actor Types of both levels (Human and Application)
*Step 2*: Counting Unadjusted Actor Weights of Type 1, which are named as Unadjusted Human Weights, briefly mentioned UHW and Unadjusted Actor Weights of Type 2 the so called Unadjusted Application Weights, briefly stated UAPW. The Total number of Unadjusted Actor Weights is (TUAW) is provided by adding up the weight values of the two Actors.
*Step3*: Compute Unadjusted Business Process Weights (UBPW)
*Step 4*: Compute Unadjusted Business Function Recovery Points (UBFRP)
*Step 5*: Define Impact Value Level and determine whether business function is included in the MBCO, by considering value of UBFRP
*Step 6*: If Function is not included in MBCO then Impact Value Level is 3 or 4 and Exercise category is *tabletop* or *medium*. Exact definition of levels and exercise categories is not important since it is considered that enterprise can survive without the specific function for a few days. However, if exact definition of the above elements is desired by the organization, the process is the same as it is in the case of complex IT functions that are included in MBCO.
*Step7*: If Function is included in MBCO then exact impact value level (1 or 2) must be defined. Determined exercise category is complex. The exact Impact Value level is calculated by considering Technical Recovery Factors, Environmental Recovery Factors and Unexpected Event Factors. Impact value level depends on the Adjusted Business Function Recovery Points (ABFRP) value and the total Recovery Effort value that will be computed. For better understanding of the analyzed algorithmic steps, a schematic presentation is also included in the current work (see Fig.1) via a UML Activity Diagram.

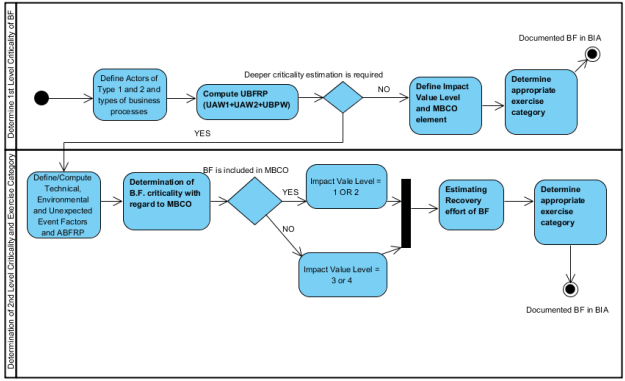

Fig. 1 Activity Diagram of the proposed BCTP approach

4.2 Definition of Business Continuity Testing Points Actor Types, Unadjusted Points and Factors

The first part of the model includes the defined mapping of Use Case Points parameters to the corresponding Business Continuity Testing Points parameters and also the calculation of the Unadjusted Points. The Actor classification of the former case is replaced by 2 different Actor Type classification in the new model. Actor Type 1 represents Human participation in the business function or process (see Table5). Moreover, Actor Type 2 represents the application involved in the same function or process (see Table6).

Table 5: Actor Type 1 BCTP Classification

| *Actor Type 1 (Human Level)* | *Weighting Factor* |
|---|---|
| Basic responsibility tasks | 1 |
| Average responsibility tasks | 2 |
| Complex responsibility tasks | 3 |

Table 6: Actor Type 2 BCTP Classification

| *Actor Type 2 (Application Level)* | *Weighting Factor* |
|---|---|
| Simple Tasks | 1 |
| Average | 2 |
| Complex | 3 |

Eq. (7) and Eq. (8) respectively provide the results of the calculation of each Actor type

$$UHW = \sum_{i=1}^{n} A1_i * W_i \qquad (7)$$

$$UAPW = \sum_{i=1}^{n} A2_i * W_i \qquad (8)$$

The Total Unadjusted Actor Weights (TUAW) is calculated by Eq. (9)

$$TUAW = UHW + UAPW \qquad (9)$$

Additionally, Use Cases are now replaced by *Business Processes*. We do not count Use Case Unadjusted weights but Business Process Unadjusted Weights. The business process complexity is now classified according to the number of process steps (see Table7). The calculation of the Unadjusted Business Process Weights is implemented with Eq. (10)

$$UBPW = \sum_{i=1}^{n} BP_i * W_i \qquad (10)$$

Table 7: Business Process Classification in BCTP

| *Business Process Type* | *Number of Process Steps* | *Weighting Factor* |
|---|---|---|
| Simple | <=3 | 1 |
| Average | 4 – 7 | 2 |
| Complex | >7 | 3 |

Finally through Eq. (11) the Unadjusted Business Function Recovery Points (UBFRP) value can be calculated,

$$UBFRP = TUAW + UBPW \qquad (11)$$

By obtaining the value of unadjusted points the first level evaluation of function criticality has been terminated. Functions that are not urgent can be simply documented to BIA template either with Impact Value Level3 and Medium Exercise Plan or with Impact Value Level4 and Tabletop Exercise plan (see Table8).

Table 8: IT Business Function Impact Value Levels and corresponding planned Exercise Category

| *Business Function* | *Impact Value Level* | *Exercise Category* | *Included in MBCO (urgent)* |
|---|---|---|---|
| BF1 | Level 1 | Complex | YES |
| BF2 | Level 2 | Complex | YES |
| BF3 | Level 3 | Medium Exercise | NO |
| BF4 | Level 4 | Tabletop Exercise | NO |

However, for urgent functions the above mapping is proposed so that disaster recovery testers will focus and test thoroughly whether the recovery time spent during the exercise meets the Rational Time Objective (RTO) of the function or its relative Maximum Acceptable Outage (MAO). In the case that none of the above hypothesis will be fulfilled the aforementioned values of the specific function should be reconsidered through a reengineering procedure. To calculate the effort spent for testing recovery plan and compare it to the RTO and MAO expected values the Technical, Environmental and Unexpected factors should be calculated as well throughout the second level of criticality estimation procedure.

### 4.2 2nd Level Business Function Criticality Estimation - Mapping Use Case Points Factors to BCTP Factors for detailed evaluation of business function

The model's second part, is currently under development. Throughout this part a deeper 2nd level estimation of function criticality is aimed via calculation of crucial factors. Environmental and Unexpected Factors, and additionally the corresponding weight of each factor are subject of future research. However the above mentioned factors have been already derived, but modified for the purposes of the new approach, from the Use Case Points model and they are listed in the following subsection (see Table9, Table10)

Table 9: Technical Recovery Factors in BCTP method

| Factor | Description |
|---|---|
| TRF1 | Application's communication with other systems |
| TRF2 | Function Type |
| TRF3 | User's skills |
| TRF4 | Complex functions |
| TRF5 | Routine functions |
| TRF6 | Easy to restore |
| TRF7 | Easy to process |
| TRF8 | Installed locally or in remote server |
| TRF9 | Exists alternative application (i.e. older) |
| TRF10 | Functional Area |
| TRF11 | Security features |
| TRF12 | Utilized by third users |
| TRF13 | Extreme and special knowledge required |

Table 10: Environmental Recovery Factors in BCTP Method

| Factor | Description |
|---|---|
| ERF1 | Familiar with Business Recovery procedures |
| ERF2 | Users' application experience |
| ERF3 | Users' recovery task knowledge |
| ERF4 | Leader's capability |
| ERF5 | Team's motivation |
| ERF6 | Stable requirements of system's recovery level (Stable MBCO) |
| ERF7 | Part – time personnel |
| ERF8 | Customers' needs direct effect |

Apart from the Technical and Environmental factors which are derived according to the Use Case Points methodology, the concept of *Unexpected Recovery Factor(s)* is introduced, enriching the state of the art in Business Continuity of IT functions (see Table11). The consideration and calculation of the specific factors is considered to be highly important; the negative influence of such factors could result to time deviation from the desired RTO and MAO of a business function recovery effort. The equation that will provide the URFactor, and the weight of each individual factor constitute subject of further research. However a short reference to each unexpected factor is required.

*Weather conditions* can constitute preventive element from reaching the recovery site or special office. The *Disaster Type* should also be considered in a testing scenario. Slight system interruption differs from building collapse in terms of recovery time. Additionally *Timely Information Distribution of Crisis Event* can result either to immediate response of staff or late response if the message is sent later on.

Table 11: Unexpected Recovery Factors in BCTP Method

| Factor | Description |
|---|---|
| URF1 | Weather conditions |
| URF2 | Disaster Type |
| URF3 | Timely Information Distribution of Crisis event |
| URF4 | Urban conditions |
| URF5 | Staff availability |
| URF6 | Network availability |

Traffic, closed roads and similar *urban conditions* can also badly affect the recovery as in the case of bad weather conditions. Moreover, *staff availability* in case of sickness and *network availability* due to technical reasons can trigger important deviation from the expected recovery time.

The final step of the BCTP model includes the calculation of the Adjusted Business Function Recovery Points (ABFRP). The value will be provided by multiplication of Unadjusted Points value, Technical Recovery Factors, Environmental Recovery Factors and Unexpected Recovery Factors according to the Eq. (12).

$$ABFRP = UBFRP * TRF * ERF * URF \qquad (12)$$

The above number will be considered towards the calculation of the Recovery Testing Effort (RTE) of a unique IT business function. The value of the effort should be less than or equal to the desired RTO, or by the worst case scenario equal to MAO. In any other case recovery tests or RTO, MAO values must be modified. The equation and method for calculating Recovery Time Effort is also a future feature of the model.

## 4. Conclusions

IT Business continuity constitutes a crucial issue of modern enterprises. The tests performed in terms of preparedness against crisis situations aim to ensure the immediate and almost continuous operation of minimum demanded IT business functions. The current work includes the proposal of a new method of planning efficient and effective recovery tests derived from the Use Case Points concept which is entitled as Business Continuity Testing Points. The accuracy of the new model is expected due to calculation of various factors that can trigger time deviation of system's recovery from the expected Recovery Time Objective or the Maximum Acceptable Outage. The current paper analyzes the initial version of the contribution. So far extension of the Use Case Points model by modifying factors, creating new actor types and introducing the Unexpected Recovery Factors has been the model's primary task. Future work and research will focus on definition of Weights of the Technical, Environmental and Unexpected Recovery Factors, development of standard mathematical equations for counting the above stated factors and creation of the final equation which will calculate the recovery effort for an individual IT business function in order to compare it with desired RTO and MAO values and finally schedule the most appropriate recovery exercise category.

**Acknowledgments**

This work was supported by ESF operational program "Education for Competitiveness" in the Czech Republic in the framework of project „Support of engineering of excellent research and development teams at the Technical University of Liberec" No. CZ.1.07/2.3.00/30.0065.

**Athanasios Podaras** received a Ph.D degree (2010) in Information Management, an MSc Diploma (2005) in Informatics from Czech University of Life Sciences in Prague, Department of Information Engineering (Czech Republic), and a Bachelor Degree in Agriculture from Technical University of Kalamata – Greece (2001). He is currently a Post - Doctoral researcher in the Technical University of Liberec, Informatics Department. His research interests include areas such as Business Process Requirement Analysis, Business Continuity Management and IT System Recovery in crisis situations, Development of Early warning Information Systems, Applied Information Technologies in Agriculture.

**Tomas Žižka** has been working since 2006 as a lecturer at the Economics Faculty of the Technical University in Liberec in the Department of Computer Science. He is currently studying a PhD study program System Engineering and Informatics in field of study called economic informatics. Title of the dissertation is the possibility of targeted receiving in non-standard situations. His research interests include areas such as Information Management in the Event of Traffic Accidents, Distribution of Information in Case of Non-standard Situations, Development of Early warning Information Systems